\documentclass{sig-alternate}

\usepackage{times}
\usepackage{graphicx}
\usepackage{epsfig}
\usepackage{url}
\usepackage{amsmath,amssymb}
\usepackage{latexsym}

\newcommand{\eat}[1]{}
\newcommand{\dbtool}[1]{{\sc Diads}}

\newcommand{\squishlist}{
  \begin{list}{$\bullet$}
   {
     \setlength{\itemsep}{0pt}
     \setlength{\parsep}{0pt}
     \setlength{\topsep}{0pt}
     \setlength{\partopsep}{0pt}
     \setlength{\leftmargin}{1.5em}
     \setlength{\labelwidth}{1em}
     \setlength{\labelsep}{0.5em} } }
\newcommand{\squishend}{
   \end{list}  }

\newcounter{ccc}

\makeatletter
\newcommand{\tinyspacing}{\let\CS=\@currsize\renewcommand
{\baselinestretch}{.9}\tiny\CS}
\newcommand{\singlespacing}{\let\CS=\@currsize\renewcommand
{\baselinestretch}{0.99}\tiny\CS}
\newcommand{\oneonespacing}{\let\CS=\@currsize\renewcommand
{\baselinestretch}{1.1}\tiny\CS}
\newcommand{\onetwospacing}{\let\CS=\@currsize\renewcommand
{\baselinestretch}{1.2}\tiny\CS}
\newcommand{\doublespacing}{\let\CS=\@currsize\renewcommand
{\baselinestretch}{1.3}\tiny\CS} \makeatother

\begin{document}

\conferenceinfo{$4^{th}$ Biennial Conference on Innovative Data Systems Research (CIDR)}{January 4-7, 2009, Asilomar, California, USA.}

\bigskip
\centerline{\bf \LARGE Why Did My Query Slow Down?\normalsize\titlenote{This work is supported by
an NSF CAREER award and faculty awards from IBM.}}
\bigskip

\centerline{\bf Nedyalko Borisov$^\dagger$ \hspace {0.25in} Shivnath Babu$^\dagger$ \hspace {0.25in} Sandeep Uttamchandani$^\ddagger$ \hspace {0.25in} Ramani Routray$^\ddagger$ \hspace {0.25in} Aameek Singh$^\ddagger$}
\smallskip
\centerline{$^\dagger$Duke University \hspace{1.5in} $^\ddagger$IBM Almaden Research Center}
\smallskip
\centerline{\{nedyalko,shivnath\}@cs.duke.edu\hspace{0.5in} \{sandeepu, routrayr, aameek.singh\}@us.ibm.com}
\bigskip

\maketitle

\begin{abstract}


Many    enterprise   environments    have    databases   running    on
network-attached  server-storage infrastructure  (referred to  as {\em
Storage Area Networks} or {\em  SANs}).  Both the database and the SAN
are complex systems that need their own separate administrative teams.
This paper puts forth the vision of an innovative management framework
to   simplify   administrative   tasks   that  require   an   in-depth
understanding  of  both the  database  and  the  SAN.  As  a  concrete
instance,  we  consider  the   task  of  diagnosing  the  slowdown  in
performance of a database query that is executed multiple times (e.g.,
in  a   periodic  report-generation  setting).   This   task  is  very
challenging  because the  space of  possible causes  includes problems
specific to the  database, problems specific to the  SAN, and problems
that arise due  to interactions between the two  systems. In addition,
the monitoring data available from these systems can be noisy.

We describe the design of
\dbtool{} which is  an integrated diagnosis tool for  database and SAN
administrators.  \dbtool{}  generates and uses  a powerful abstraction
called  {\em Annotated  Plan  Graphs (APGs)}  that  ties together  the
execution  path of  queries in  the database  and the  SAN.   Using an
innovative  workflow  that  combines  domain-specific  knowledge  with
machine-learning  techniques,~\dbtool{}  was  applied successfully  to
diagnose  query slowdowns  caused  by complex  combinations of  events
across a PostgreSQL database and a production SAN.


\eat{

We present {\sc DiaDS}, an integrated DIAgnosis tool for Databases and Storage area networks
(SANs). Existing diagnosis tools in this domain have a database-only (e.g., ~\cite{tune-oracle})
or SAN-only (e.g., ~\cite{shen05perf}) focus. 
\dbtool{} is a first-of-a-kind framework
based on a careful integration of information 
from the database and SAN layers; and is not a simple concatenation of 
database-only and SAN-only modules. This approach 
not only increases the accuracy of diagnosis, but also leads to significant improvements in efficiency.

\dbtool{} uses a novel combination of non-intrusive machine learning techniques (e.g., Kernel Density Estimation)
and domain knowledge encoded in a new symptoms database design. The machine learning part
provides core techniques for problem diagnosis from monitoring data, and domain knowledge 
acts as checks-and-balances to guide the diagnosis in the right direction. This unique system design enables
\dbtool{} to function effectively even in the 
presence of multiple concurrent problems as well as 
noisy data prevalent in production environments. 
We demonstrate the efficacy of our approach 
through a detailed experimental evaluation of~\dbtool{} implemented on a real data center testbed with PostgreSQL databases and an enterprise SAN.

} 


\end{abstract}

\section{Introduction}
\label{sec:intro}

Database deployments in enterprise environments are typically business
critical and support high transaction rates.  These deployments run on
enterprise-class storage subsystems with terabyte-scale data mapped to
the database either through a  file system (referred to as {\em System
Managed Storage}) or raw volumes (referred to as {\em Database Managed
Storage}).  Traditionally,  storage was attached  directly to high-end
database  servers to  meet their  capacity, throughput,  and bandwidth
requirements.  However, economic realities of high administration costs
for islands of disconnected resources, combined with under-utilization
of   statically-provisioned   server   and  storage   hardware,   have
transformed the direct-attached  architectures into a network-attached
setup   with  multiple   application  servers   (including  databases)
connected  to   a  consolidated  and  virtualized   storage  pool;  an
architecture known popularly as a {\em Storage Area Network (SAN)}.

SANs are very complex systems.  A  typical SAN has a hierarchy of {\em
core}  and  {\em  edge}   fibre-channel  switches  with  {\em  zoning}
configuration that controls the  connectivity of server ports with one
or  more heterogeneous  storage controllers.   The  storage controllers
manage a  large number of raw  disks by aggregating  them into logical
entities  like pools  and  volumes.  Given  this complexity,  database
administrators are forced to treat  the SAN as a black-box, entrusting
SAN administrators to configure the required CPU, network, and storage
resources for meeting their database's performance requirements.

Such a  {\em silo-based} approach  for database and SAN  management is
the state-of-art  today.  In  a typical real-world  scenario, database
administrators  open  problem tickets  for  the  SAN administrator  to
analyze and fix  issues related to query slowdowns:  {\em ``Queries to
the RepDB database used for report generation have a 30\% slow down in
response time, compared to performance two weeks back.''} Unless there
is an  obvious failure or degradation  in the storage  hardware or the
connectivity fabric, the SAN  administrator's response to this problem
ticket could be: {\em ``The  I/O rate for RepDB tablespace volumes has
increased 40\%, with increased sequential reads, but the response time
is  within normal  bounds.''}  This  ``blame game''  may  continue for
several weeks before  the problem is actually fixed.   In reality, the
query slowdown problem could be  due to any number of causes including
suboptimal  plan  selection by  the  database  due  to incorrect  cost
models, lock contention  for the database tables, CPU  saturation of a
database server, congestion in  the controller ports, and others.  The
lack  of consistent  end-to-end information  may lead  to  either {\em
throwing iron  at the problem}  and creating islands  of underutilized
resources, or  employing highly  paid consultants who  understand both
databases and SANs to solve the original problem tickets.

Our vision in this paper  is an integrated database and SAN management
framework. This framework combines details of both database operations
as  well  as SAN  configuration  and  performance  into a  novel  data
structure referred  to as  an {\em Annotated  Plan Graph  (APG)}.  The
framework uses a combination of machine learning algorithms and domain
knowledge  to help administrators  with key  day-to-day tasks  such as
optimized allocation  of SAN  resources for varying  database workload
characteristics,  diagnosis  of  database performance  slowdowns,  and
what-if analysis  related to workload or configuration  changes.  As a
concrete  instance of our  vision, this  paper focuses  on integrated
diagnosis  of query  performance  slowdown in  databases running  over
SANs.


\subsection{Challenges in Integrated Diagnosis}

Enterprise environments  are constantly  evolving with changes  in the
SAN  configuration,  the mix  of  database  queries,  as well  as  the
workload  characteristics of  other applications  sharing the  SAN. In
such an environment, the key challenges for diagnosis are as follows:

\vspace{1mm}

\squishlist
\item  {\em Cascading  of events}:  Analyzing the  impact of  an event
across multiple layers of a  system is a nontrivial problem. The cause
and effect  of a problem may  not be contained within  a single layer,
but manifested  across multiple layers (typically referred  to as {\em
event flooding}).

\vspace{1mm}

 \item {\em Inaccuracies in monitoring data}: Monitoring in production
environments is  configured to minimize  the impact on  the foreground
applications.   Typically,  the  monitoring  intervals  are  large  (5
minutes or  higher), which  may lead to  inaccuracies (referred  to as
{\em  noisy} data)  because the  instantaneous effects  of  spikes and
other bursty behavior can get averaged out.

\vspace{1mm}

\item {\em  High dimensional search space  with complex correlations}:
An integrated  analysis involves a large number  of entities including
database operators, physical SAN devices, logical volumes and pools in
a SAN, and workload. Pure machine learning techniques that aim to find
correlations or regression functions in the raw monitoring data, which
otherwise  may have  been  effective  within a  single  layer, can  be
ineffective in the integrated  scenario.  Existing diagnosis tools for
some commercial databases~\cite{tune-oracle} use a rule-based approach
where  a root-cause  taxonomy is  created and  then  complemented with
rules to  map observed symptoms  to possible root causes.   While this
approach  has the  merit  of encoding  valuable  domain knowledge  for
diagnosis purposes, it may become complex to maintain and customize.

\squishend


\subsection{Contributions}

Our vision is  to leverage the existing monitoring  tools for SANs and
databases  to  develop  an  integrated  database  and  SAN  management
platform.  This  platform will  simplify the subset  of administrative
tasks that require an understanding  of both databases and SANs, e.g.,
problem  diagnosis,  resource   provisioning,  what-if  analysis,  and
disaster recovery planning.  As  a concrete instance of the integrated
functionality,  the paper  describes  our prototype  of an  integrated
diagnosis tool (referred to  as~\dbtool{}) that spans the database and
the underlying SAN that consists of end-to-end I/O paths with servers,
interconnecting network switches  and fabric, and storage controllers.
Figure~\ref{fig:apgs}  shows an integrated  database and  SAN taxonomy
with  various logical  (e.g., sort  and scan  operators)  and physical
components (e.g., server, switch, and storage subsystem).

To the  best of our  knowledge,~\dbtool{} is the first  diagnosis tool
that analyzes both  SAN and database events in  an integrated fashion.
The key contributions of this paper are:

\vspace{1mm}

\squishlist

\item A  novel canonical  representation of database  query operations
combined with  physical and logical entities from  the SAN environment
(referred  to as  {\em  Annotated Plan  Graphs}). This  representation
captures  the information  required for  end-to-end diagnosis,  and is
created using monitoring data from available database and SAN tools.

\vspace{1mm}

\item  An  innovative  diagnosis   workflow  that  {\em  drills  down}
progressively from  the level  of the query  to database plans  and to
operators, and  then uses  configuration dependency analysis  and {\em
symptom signatures} to further drill  down to the level of performance
metrics and events in components.  It then {\em rolls up} using impact
analysis  to tie potential  root causes  back to  their impact  on the
query slowdown.  The diagnosis  is accomplished using a combination of
machine learning and domain knowledge

\vspace{1mm}

\item  An empirical  evaluation of~\dbtool{}  on a  real-world testbed
with  a PostgreSQL  database  running on  an enterprise-class  storage
controller. We describe  (and demonstrate) problem injection scenarios
including combinations of events at the database and SAN layers, along
with  a  drill-down   into  intermediate  internal  results  generated
by~\dbtool{}.

\squishend

\section{Related Work} \label{section:related}

There  has  been much  prior  research  for  performance diagnosis  in
databases~\cite{tune-oracle,automatic-hp}   as   well  as   enterprise
storage  systems~\cite{genesis,shen05perf}.  However,  most  of  these
techniques  perform  diagnosis in  an  isolated  manner attempting  to
identify root cause(s) of a performance problem in individual database
or storage silos. Since the performance  problem may lie in any one or
a combination  of database (DB)  and SAN layers, an  integrated system
like \dbtool{} would be a useful and more efficient approach.

Recent  studies  that  have  looked  at  the  interdependence  between
database  and storage  systems  highlight the  importance  of such  an
integrated analysis.   Reference \cite{reiss-sigmod} described  how an
inaccurate  storage cost  model in  the database  query  optimizer can
significantly impact  the choice of query  execution plans.  Reference
\cite{salem}  proposed  an end-to-end  database  and storage  planning
technique  by  characterizing the  storage  I/O  workload  of a  given
database  workload using  an independent  combination of  database and
storage analysis.   While sharing the  same spirit, our work  brings a
much tighter coupling  of database-level and storage-level information
as  well as capturing  their interdependence  using a  novel Annotated
Plan Graph abstraction described in Section \ref{sec:apgs}.

\dbtool{} can be a  good complement to fine-grained database diagnosis
and tuning  tools like Oracle's Automatic  Database Diagnostic Monitor
(ADDM)~\cite{tune-oracle}.  ADDM is a database profiling and diagnosis
tool  that  uses  expert  knowledge  about the  database  to  identify
problems  as  well  to  recommend  possible  fixes  to  the  problems.
Reference \cite{sqlcm} describes a server-side monitoring and analysis
system  for  Microsoft  SQL   Server  that  is  useful  during  manual
diagnosis.   Our  work  complements   this  research  by  providing  a
non-intrusive  and low-overhead  mode of  analysis that  uses historic
performance data  to diagnose {\it changes} in  query performance.  We
discuss this synergy further in Section~\ref{sec:potential}.

There  has  also  been  significant  work  in  diagnosing  performance
problems  within  the  systems research  community~\cite{peerpressure,
symptom-db}.   Broadly,  these  techniques   can  be  split  into  two
categories:  (a) systems  using machine  learning techniques,  and (b)
systems using domain knowledge. Reference \cite{peerpressure, pc-slow}
uses statistical  techniques to develop models for  a healthy machine,
and uses  the models  to identify {\it  sick} machines.  On  the other
hand,  systems  like~\cite{codebook,  symptom-db, symptom-format, chilukuri06}  use
domain knowledge  to create a {\it symptoms}  database that associates
performance symptoms with underlying  root causes.  Such databases are
often  created manually  and require  a  high level  of expertise  and
resources to maintain.

We believe that  for a diagnosis tool to be  practically useful, a mix
of  machine learning  and  domain knowledge  will  be required.   Pure
machine learning techniques can be misled due to spurious correlations
in data resulting from noisy  data collection or event flooding (where
a   problem  in  one   component  causes   another  component   to  be
impacted). In \dbtool{}, we  counterbalance this effect using suitable
domain knowledge like  component dependencies, symptoms databases, and
knowledge of query plan and operator relationships.

Next, we describe Annotated Plan Graphs that capture database and storage component behavior in a single integrated abstraction.

\section{Annotated Plan Graphs}
\label{sec:apgs}

Suppose a query $Q$ that a report-generation application issues periodically to the database system shows a slowdown in
performance.  The root cause of this slowdown may lie in the database layer (execution plan becoming suboptimal due to
changes in data properties) or the SAN layer (increased congestion in the storage pool) or often a combination of the
two. Diagnosing such a problem requires the ability to understand the behavior of not only the database and storage
layers during the execution of the query, but also the interaction between the two layers.

\begin{figure}[t!]
\centering 
\includegraphics[clip=true,width=0.6\linewidth,height=7in,viewport=0 120 245 700]{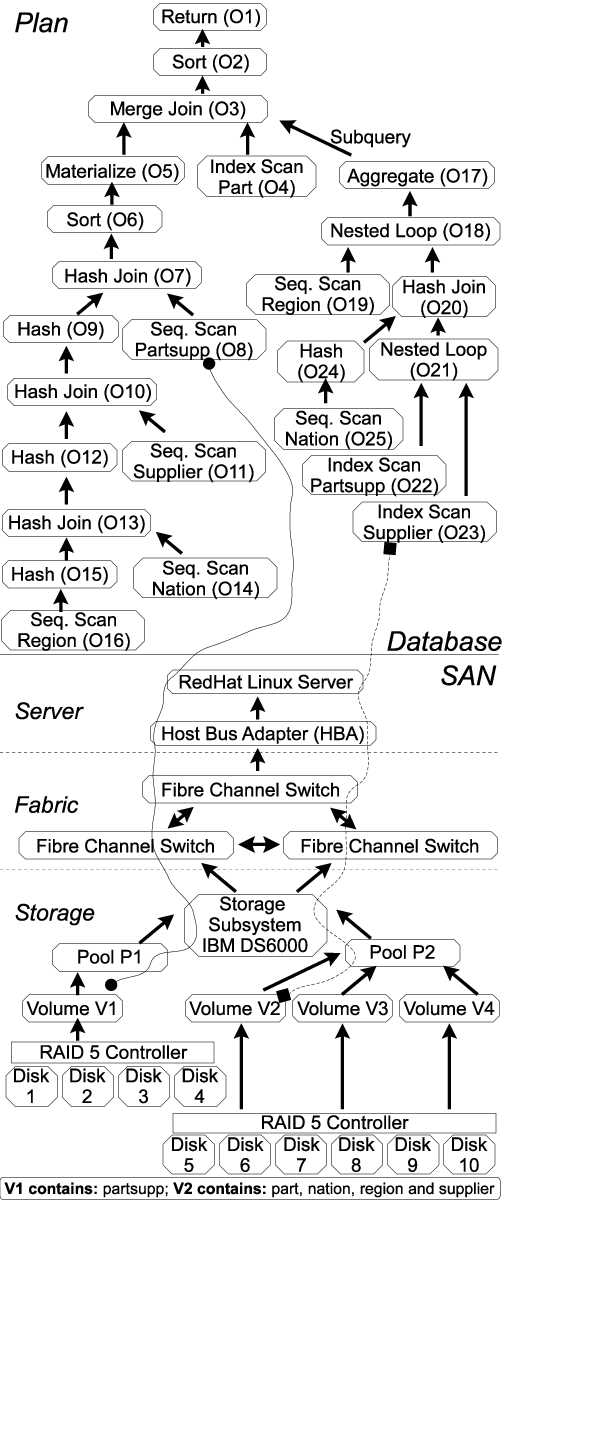}
\vspace{-3mm}
\caption{Annotated Plan Graph}
\label{fig:apgs}
\vspace{-5mm}
\end{figure}

The  Annotated  Plan Graph  (APG)  abstraction  provides this  precise
ability. At a  high level, an APG captures  a comprehensive end-to-end
mapping of  the logical  database operators of  the query plan  to the
physical disk details where the actual data resides, and everything in
between.  Figure  \ref{fig:apgs} shows an  example of an  APG instance
for Query 2 from TPC-H. In  the database layer, APG includes the query
plan consisting  of 25 operators, denoted $O_1$-$O_{25}$,  with 9 leaf
operators.   Below  the database  layer,  APG  also  includes the  SAN
configuration  and  correlations  including servers,  storage  network
fabric,  storage  pools ($P_1$,  $P_2$)  and  storage volumes  ($V_1$,
$V_2$) containing the database tables.

We   describe   the   complete   process  of   APG   construction   in
Section~\ref{sec:apg-construct}.  As a  quick summary, an APG contains
the following  kinds of  information.  At the  database level,  an APG
includes:

\squishlist

\vspace{1mm}

\item {\em Query-level data:} For each execution of plan $P$, \dbtool{} collects some low-overhead monitoring data per
operator $O \in P$. The relevant data includes: $O$'s start time, stop time, and {\em record-counts} (estimated and
actual number of records in $O$'s output in the plan).

\vspace{1mm}

\item  {\em  Database-level data}  includes  common  metrics like  the
number of buffer cache hits,  full-table scans, random I/Os, and locks
held.

\squishend 

\vspace{1mm}

The data  collected at  the SAN level  includes: (i)  configuration of
components  (both  physical  and  logical),  (ii)  connectivity  among
components,   (iii)   changes   in  configuration   and   connectivity
information  over time,  (iv) performance  metrics from components, (v)
events generated by the system  (e.g., disk failure, RAID rebuild) and
(vi)   events  generated   by  user-defined   {\em   triggers}  (e.g.,
degradation   in  volume   performance,  high   workload   on  storage
subsystem).

It is important to note that the APG abstraction is first of a kind in
this area. Several product  offerings (e.g., \cite{hp-sim,tpc}) in the
market today,  while collecting monitoring data from  IT systems, only
contain  silo-based  database   or  SAN  information.   These  product
offerings cannot trace the flow of requests across multiple subsystems
either because  such tracing is  impossible (e.g., the  subsystems are
from multiple vendors) or it  is impractical (e.g., the load placed on
the  production system  is high).   Thus, no  current tool  provides a
convenient abstraction that captures  query behavior in a seamless way
across the  database and the  SAN.  Using low-overheard  monitoring of
historic performance data, an APG fulfills this need.

Some of the novel features of APGs are:

\squishlist

\vspace{1mm}

\item APGs are generated from light-weight monitoring data that is readily available in most production environments. 

\vspace{1mm}

\item APGs are views on the monitoring data that combine what DBAs see---e.g., data on query plans---with what SAN
administrators see---data from the numerous SAN components and their interconnections. More importantly, APGs show each
administrator what she  typically does not get to see.  However, APGs are much more than a juxtaposition of these two
pieces of data; as discussed next. 

\vspace{1mm}

\item APGs capture the dependency paths of their constituent components.  For example, the dependency path of an
operator $O$ is the set of physical (e.g., CPU, database cache, disk) and logical (e.g., volume, workload) system
components whose performance can impact $O$'s performance. There are {\em inner} and {\em outer} dependency paths. The
performance of components in $O$'s inner dependency path can affect $O$'s performance directly.  $O$'s outer dependency
path consists of components that affect $O$'s performance indirectly by affecting the performance of components on the
inner dependency path.  As an example, the inner dependency path for the Index Scan operator $O_{23}$ in  Figure
\ref{fig:apgs} includes the server, HBA, FCSwitches, storage subsystem, Pool $P_2$, Volume $V_2$, and Disks $5$-$10$.
The outer dependency path includes Volumes $V_3$ and $V_4$ (because of the shared disks) and other database queries.
Section \ref{sec:design} discusses how these dependency paths can be {\em pruned} using correlation analysis. 

\vspace{1mm}

\item Each component in an APG is {\em annotated} with appropriate monitoring data collected during the plan's
execution. For example, the annotation of an operator $O$ consists of the performance data   collected by \dbtool{} for
each component $C$ in $O$'s dependency path; this data is collected in the $[t_b,t_e]$ time interval where $t_b$ and
$t_e$ are respectively $O$'s (absolute) start and stop times for that execution.

\squishend 

\vspace{-1mm}

\subsection{Construction of APGs}
\label{sec:apg-construct}

Constructing an APG involves a number of steps including gathering configuration and performance of SAN and the database
components. In this section, we describe how \dbtool{} collects and correlates this data starting with the SAN component
of the APG.

\vspace{-1mm}

\subsubsection{SAN Layer Data}

\noindent  {\bf Configuration Data}:  Enterprise SANs  are based  on a
shared storage  model with multiple servers (DB  servers, file servers,
etc.) accessing  the same backend  storage. The storage  media (disks)
are contained  within storage subsystems, also referred  to as storage
controllers, which  provide specialized functionality  like RAID, copy
services, and  storage virtualization.   The servers are  connected to
these storage  subsystems through a  hierarchy of switches  usually as
part     of     a     Fibre     Channel    (FC)     network     fabric
(Figure~\ref{fig:apgs}). Each server has one or more Host Bus Adapters
(HBAs) with  one or more FC  ports each. These ports  are connected to
multiple layers of  SAN switches which are then  connected to ports on
storage subsystems.

Within a subsystem, storage is abstracted through logical storage {\it
pools} which are carved into  storage {\it volumes} that are then made
accessible to desired servers.  Data contained within these volumes is
physically placed onto  the disks that comprise the  logical pool, and
may  be striped  across these  disks  depending upon  the chosen  RAID
configuration.   Additionally,  two  important configuration  settings
dictate the accessibility of data to servers: (i) {\it Zoning} dictates
which storage subsystem ports can be accessed by any given server, and
(ii)  {\it Logical  Unit Number  (LUN)  mapping/masking} configuration
defines  the storage  volumes on  the  storage subsystem  that can  be
accessed  by   a  particular  host.   Referring  to   the  example  in
Figure~\ref{fig:apgs}, the database instance  is installed on a Redhat
Linux Server  connected to  a Fibre Channel  Switch, and  uses storage
from  an IBM  DS6000  Storage controller.  $O_{23}$  operates on  data
residing  in  Volume $V2$  which  is part  of  the  storage pool  $P2$
(illustrated by a dotted line in the figure).

To manage these  complex interconnected and interdependent components,
administrators   often   use  storage   management   tools  like   IBM
TPC~\cite{tpc}  or  EMC  Control  Center~\cite{emc-cc}.   These  tools
collect the  entire configuration information  from the SAN  and store
them  in   a  centralized  database.    In  recent  years,   both  the
communication  with  SAN  devices  as  well  as  the  schema  for  the
configuration  database are  available as  open standards,  e.g., SNIA
SMI-S~\cite{smis}    for   device    configuration    collection   and
Aperi~\cite{aperi}   for   the   configuration   database.   For   our
implementation of \dbtool, we use  the IBM TPC database to extract SAN
configuration information  which is then correlated  with the database
layer    information   to    construct   the    APG   as    shown   in
Figure~\ref{fig:apgs}.\\

\vspace{-2mm}

\noindent {\bf  Performance Data}:  Standards like SMI-S  also include
performance data  for various  logical and physical  components (e.g.,
servers, HBAs, storage pools,  and storage volumes).  This performance
data is  critical for  \dbtool{} analysis.  For  instance, performance
attributes monitored from a server and its HBA include {\em Percentage
CPU Usage,  Free Memory, Cache  Hit Rate, Process Start  Time, Process
End Time  CPU, Percentage Utilization  by process, Memory  Consumed by
Process,  IO Count,  IO  per  second, HBA  FC  Port Statistics}.   For
storage components,  \dbtool{} uses  attributes like {\em  Bytes Read,
Bytes  Written,  Sequential   Read  Hits,  Sequential  Read  Requests,
Sequential Write Requests,  Total IOs}. An anomalous value  for any of
these attributes during a query slowdown is a candidate for additional
evaluation in~\dbtool{}. This performance  data is used as annotations
in the APG for each respective component.

\eat{This historic performance data can be seen as a time series
of APGs corresponding to the query $Q$ of interest. The diagnosis problem is to relate $Q$'s slowdown to what change in
the rest of the system caused the slowdown.}

\vspace{-1mm}

\subsubsection{Database Layer Data and Correlation with SAN Data}
In the  database layer,  an APG includes  the query  plan information,
including  data from  all the  operators in  the plan.   This  data is
easily  available   in  most  DBMSs.   (We  used   PostgreSQL  in  our
implementation  of   the  \dbtool{}  prototype.)    Each  operator  is
annotated with  the performance data  of components in  its dependency
paths.

The dependencies between operators  and SAN components are obtained in
the  following manner.   It begins  with the  parsing of  the database
configuration  file   that  defines   the  mapping  of   the  database
tablespaces  to  the  storage  volumes  in the  SAN.   There  are  two
predominant  configurations  for  associating  physical storage  to  a
tablespace defined by the  database: (a) System Managed Storage (SMS),
where  the tablespace  is mapped  to a  file system  created on  a SAN
volume; (b)  Database Managed Storage  (DMS), where the  tablespace is
created  on  a raw  physical  SAN  volume  with space  allocation  and
associated  book-keeping managed directly  by the  database. Operators
$O_1$ to $O_{25}$ in Figure \ref{fig:apgs} are related either directly
or indirectly  to operations on  tables, which belong  to tablespaces.
Thus, given  an operator  $O$, it is  possible to  map $O$ to  the SAN
volumes  that $O$  depends on.   Combining this  mapping with  the SAN
configuration data, we can obtain the inner and outer dependency paths
for all plan operators.

\eat{With respect to collecting database-level and query-level stats, most enterprise databases expose a rich set
of APIs to collect such details. The IO path information from the database server to storage volumes is available from
SAN monitoring tools. For the implementation of APG, \dbtool{} uses the IBM TotalStorage Productivity Center \cite{tpc}
information related to the SAN. The collected data is transformed into a tabular format, and persisted as time-series
data in an internal database.}


\section{Design of {\large \dbtool{}}}
\label{sec:design}

\begin{figure}[t]
\centerline{\epsfxsize=3.5in \epsfysize=3.7in
\epsffile{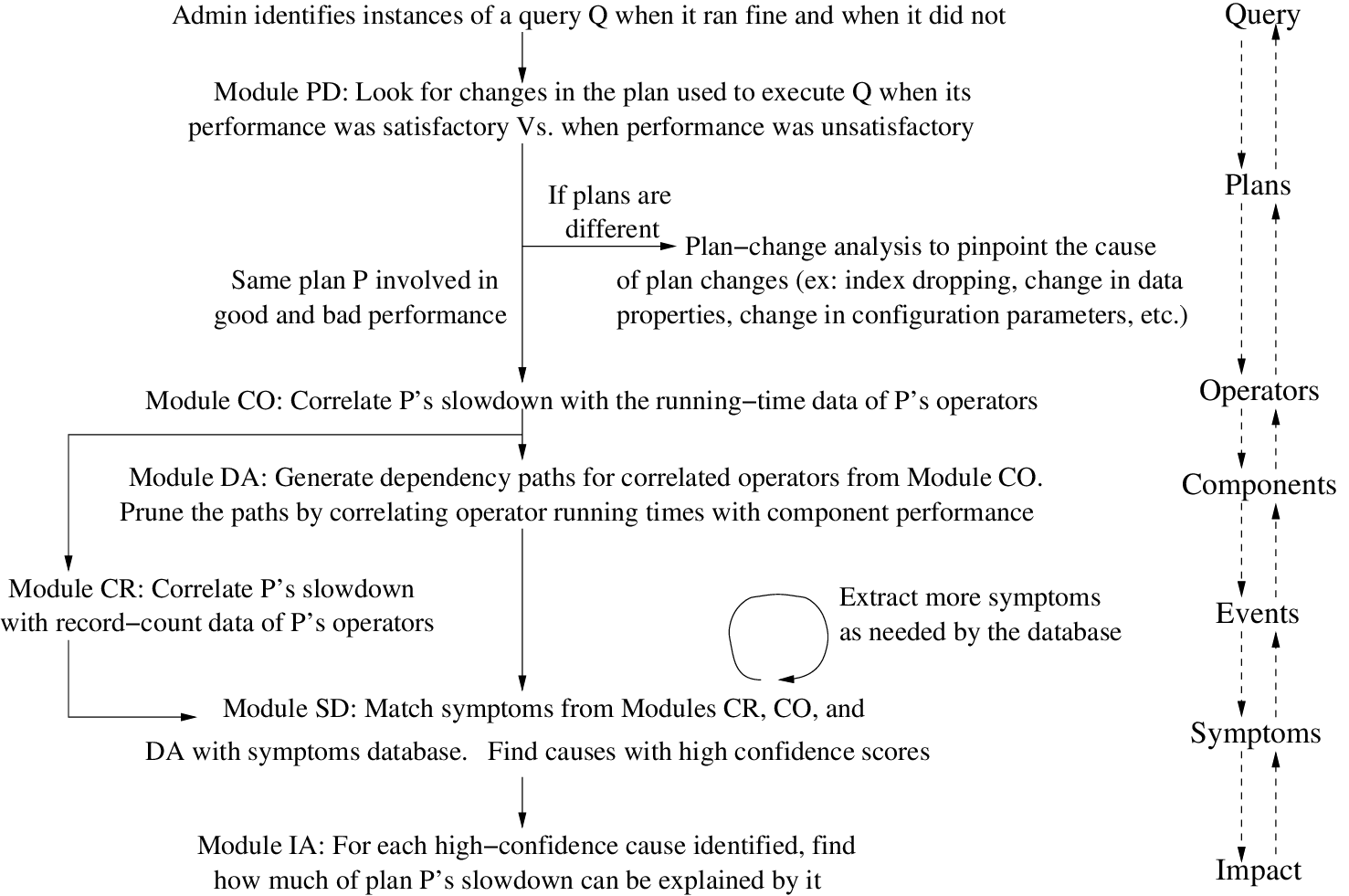}}
\vspace {-0.5em}
{\caption{\dbtool{}'s diagnosis workflow}
\label{fig:workflow}}
\protect \vspace{-5mm}
\end{figure}

When the administrator identifies a query $Q$ as having experienced a slowdown, 
\dbtool{} invokes the {\em diagnosis workflow} shown in Figure \ref{fig:workflow}.
This workflow ``drills down" progressively from the level of the query to plans
and to operators, and then further down to the level of performance metrics and events in components. 
Finally, an impact analysis is done that ``rolls up" to tie potential
root causes back to their impact on $Q$'s slowdown. As we will show in this section, 
the workflow applies a combination of statistical machine learning and domain knowledge to the 
APGs collected for $Q$. This novel combination provides built-in checks and balances to deal
with the challenges listed in Section \ref{sec:intro}.

\subsection{Modules in the Diagnosis Workflow}
\label{sec:design_modules}

The administrator first specifies declaratively or marks directly the runs of the
query that were {\em satisfactory} and those that were {\em unsatisfactory}. 
For example, runs with running time below 100 seconds may be satisfactory, or 
all runs from 8 AM to 2 PM were satisfactory, and those from 2 PM to 3 PM were unsatisfactory.

\eat{
\textbf{Module Plan Diffing (PD):}

In the case when historic satisfactory data for the run of plan $P$ is not present, 
this module provides reasoning for the database decision of choosing plan 
$P$ for the execution of the query. This reasoning is based on 
the structure comparison between plan $P$ and the the query's most recent executed 
plan (defined as $P'$) that had satisfactory run. Furthermore, the tool process 
all configuration changes that have been observed between the runs of $P'$ and $P$, and 
using the database cost models assigns impact score to each configuration change. Finally,
the impact scores are used as ranking to provide the administrator information of the
changes that could have led to the execution of plan P. 

As future work we envision improvements on this scheme with more accurate and precise
methods for plan difference reasoning.
}

\textbf{Module Plan Diffing (PD):} 
	The first module in the workflow looks for significant changes 
	between the plans used in satisfactory and unsatisfactory runs. 
	If such changes exist---e.g., if \dbtool{} finds that plan $P_1$ was used 
	in satisfactory runs and a different plan $P_2$ was used 
	in unsatisfactory runs---then \dbtool{} tries to pinpoint the cause of the
	plan changes (which includes, e.g., index addition or dropping, 
	changes in data properties, or changes in configuration parameters used during plan selection).
Our current implementation considers each 
schema or configuration change that occurred between the runs of $P_1$ and $P_2$, and 
checks whether this change could have caused the plan change.
	The remaining modules in the workflow are invoked if \dbtool{} finds a plan $P$ that is involved in both
	satisfactory and unsatisfactory runs of the query.

\textbf{Module Correlated Operators (CO):} 
The objective of this module is to find the subset of operators, called the {\em correlated operator set (COS)},
whose change in performance best explains plan $P$'s slowdown. COS is identified by 
analyzing data from satisfactory and unsatisfactory runs of $P$ which can be 
seen as records with attributes $L,t(P),t(O_1),t(O_2),\ldots,t(O_n)$ for each run of $P$. (Recall the annotations
maintained for APGs in Section \ref{sec:apgs}.) Here, attribute $t(P)$ is 
the total time for one complete run of $P$, and attribute
$t(O_i)$ is the running time of operator $O_i \in P$ for that run.
Attribute $L$ is a {\em label} representing whether the corresponding run of $P$ was satisfactory or not.

\dbtool{} feeds this data to \textit{Kernel Density Estimation (KDE)} which 
is a statistical method to estimate the probability density function of a random variable.
KDE applies an estimator to the data to learn the probability density function $f_i(S_i)$ of
the random variable $S_i$ representing the running time of operator $O_i$ when $P$'s performance is satisfactory. 
Let $u$ be an observation of $O_i$'s running time when 
$P$'s performance was unsatisfactory. Consider the probability estimate 
$\mbox{\em prob}(S_i \leq u) = $ $\int_{-\infty}^{u} f_i(S_i) ds_i$. Intuitively, 
as $u$ becomes higher than the typical range of values of $S_i$, 
$\mbox{\em prob}(S_i \leq u)$ becomes closer to 1. Thus, a high value of $\mbox{\em prob}(S_i \leq u)$
represents a significant increase in $O_i$'s running time when plan performance was
unsatisfactory; if so, $O_i$ belongs to COS.
$\mbox{\em prob}(S_i \leq \overline{u})$ is called the {\em anomaly score} of operator $O_i$.

\textbf{Module Dependency Analysis (DA):} 
This module identifies the subset of system components, called the {\em correlated component set (CCS)},
such that each component in CCS: (i) is in the dependency path of at least one operator $O \in $ COS, and
(ii) has at least one performance metric that is significantly correlated with $O$'s running time. 
The fact that a component $C$ is in the dependency path of an operator $O \in $ COS (Property (i) above)
does not necessarily mean that $O$'s performance has been affected
by $C$'s performance. Hence, \dbtool{} checks additionally for Property (ii) which is implemented 
as correlation analysis using KDE.

\textbf{Module Correlated Record-counts (CR):} 
In this module, \dbtool{} checks whether the change in performance of operators in COS
correlates with their record-counts. Significant correlations mean that data properties have changed
between satisfactory and unsatisfactory runs of $P$. Once again, correlation analysis
is implemented using KDE to find  the {\em correlated record-count set \textit{CRS}} $\subseteq$ COS.

\textbf{Module Symptoms Database (SD):} COS, CCS, and CRS along with other observed SAN and database events may only be \textit{symptoms} of the true root cause of $P$'s slowdown. Module SD seeks to map the observed symptoms to the actual root cause. \dbtool{} generates this mapping using a {\em symptoms database} whose main purpose
is to streamline the use of domain knowledge to (i) create more accurate results by dealing with event propagation, and (ii) generate semantically meaningful results (e.g., reporting lock contention as a cause 
instead of reporting some performance metrics only). \dbtool{}'s implementation of the symptoms database
is motivated by an intuitive and commercially-used format called the {\em Codebook}. The original format assumes
a finite set of symptoms such that each distinct root cause $R$ has a unique {\em signature} in this set.
However, \dbtool{} needs to consider complex symptoms such as symptoms with temporal properties
(e.g., contention occurred before failure).

 \dbtool{}'s symptoms database is a collection of root cause entries each of which 
has the format {\em Cond}$_1$ $\&$ {\em Cond}$_2$ $\&$ $\ldots$ $\&$ {\em Cond}$_z$, 
for some $z > 0$ which can differ across entries. Each {\em Cond}$_i$ is a condition of the form 
$\exists \mbox{\em symp}_j$ (denoting presence of $\mbox{\em symp}_j$)
or $\neg \exists \mbox{\em symp}_j$ (denoting absence of $\mbox{\em symp}_j$).
Symptom $\mbox{\em symp}_j$ is represented in a high-level language
used to express complex symptoms over a base set of symptoms \cite{act}.
Each {\em Cond}$_i$ is associated with a weight $w_i$ such the sum of the weights
for each individual root cause entry is 100\%.
From the symptoms observed currently, \dbtool{} calculates a {\em confidence score} for each root cause
$R$ as the sum of the weights of $R$'s conditions that evaluate to true. We further divide the confidence score into
three categories: (i) high (score $\geq 80\%$), (ii) medium ($80\% >$ score $\geq 50\%$), and (iii) low 
(score $< 50\%$).

\textbf{Module Impact Analysis (IA):}
For each high-confidence root cause $R$ identified by Module SD, an \textit{impact score} is calculated
as the percentage of the query slowdown (time) that can be contributed to $R$ individually.
When multiple problems coexist in the system, impact scores can separate out high-impact
causes from the less significant ones. Also, they serve 
as a safeguard against misdiagnoses resulting from spurious correlations due to noise.

\dbtool{} has multiple implementations of this module. One implementation is an  
``inverse dependency analysis". First, IA starts from a root cause ($R$) and identifies all system components affected by $R$, denoted $comp(R)$. The next step is to find the subset of operators ($op(R)$) whose performance
is affected by $comp(R)$. The impact score is calculated as the percentage of extra running time of $op(R)$ 
with respect to the extra plan running time; where extra time is the difference 
between the average running times across unsatisfactory and satisfactory runs. 
Another implementation of IA leverages the plan cost models used by database query optimizers.

\begin{table*}[t]
	\centering  
	\begin{tabular} {|p{4.2in}|p{2.5in}|} 
		\hline 
		\textbf{Problem Description} & \textbf{Critical Role of \dbtool{} Modules in Diagnosis} \\ 
		\hline
		1. SAN misconfiguration leading to contention in volume $V_1$ & Identified symptoms pinpoint the correct volume; SD maps symptoms to the correct root cause \\ \hline 
		2. Contention caused by external workloads on volumes $V_1$ and $V_2$; with only the former affecting query performance & DA prunes out the unrelated symptoms and events for volume V2 \\ \hline
		3. SQL DML causes a subtle change in data properties; problem propagates to SAN causing volume contention & CR identifies the important symptoms; IA rules out volume contention as a root cause  \\ \hline 
		4. Concurrent DB (change in data properties) and SAN (misconfiguration) problems & Both problems identified; IA correctly ranks them\\ \hline 
		5. DB problem (locking-based) and spurious symptoms of volume contention due to noise & IA identifies volume contention as low impact \\ \hline
	\end{tabular}
	\vspace{-4mm}
	\caption{Experimental settings of increasing complexity used to evaluate \dbtool{}} 
	\label{table:exper}
	\vspace{-2mm} 
\end{table*}

\section{Experimental Evaluation}
\label{sec:exper}

For the evaluation of \dbtool{}, we considered query slowdowns 
caused by problems within the database and SAN layers as
well as combinations of problems across both layers (a capability which
is unique to~\dbtool{}). Our experimental testbed is part of a production SAN environment, 
with the interconnecting fabric and storage controllers being shared by other applications.
The testbed runs TPC-H queries on a PostgreSQL database server 
configured to access tables using two Ext3 file system volumes $V_1$ and $V_2$ created
on an enterprise-class storage controller. Figure \ref{fig:apgs} shows the table layout
and the query plan that we will focus on.

Table \ref{table:exper} gives a high-level summary of our experimental
scenarios. \dbtool{} successfully diagnosed the root cause in all these 
cases. In this paper we will focus only on the first scenario; 
additional analysis for all scenarios are covered in~\cite{diads}. 
In the first scenario, a
contention is created in volume $V_1$ (from Figure \ref{fig:apgs})
causing a slowdown in query performance. The root cause of the
contention is another application workload that is configured in the
SAN to use a volume V' that gets mapped to the same physical disks as
$V_1$. For an accurate diagnosis result, \dbtool{} needs to pinpoint
the combination of SAN configuration events generated on: (i) creation
of the new volume V', and (ii) creation of a new zoning and mapping
relationship of the server running the workload that accesses V'.

\noindent {\bf Modules PD and CR:} These two modules correctly identify (respectively) that the plan 
and the data properties have not changed.

\noindent {\bf Module CO:} Based on KDE, this module identifies the set of correlated operators as $O_2$, $O_3$, $O_4$, $O_6$, $O_7$, $O_8$, $O_{17}$, $O_{18}$, $O_{20}$, $O_{21}$ and $O_{22}$ (each operator has an anomaly score greater than the threshold of 0.8). This set correctly contains both the leaf operators ($O_8$ and $O_{22}$) connected to volume $V_1$. The eight intermediate operators present in this set are ranked highly 
because of event propagation: the running times of these 
operators are affected by the running times of the ``upstream" operators (in this case $O_8$ and $O_{22}$).
Finally, operator $O_4$ is a false positive because it operates on volume $V_2$, and is not affected
by the contention in $V_1$. (As we will see shortly, this false positive caused by noise gets filtered out.)

\noindent {\bf Module DA:} This module computes anomaly scores for performance metrics in both volumes 
$V_1$ and $V_2$ since these volumes fall in the dependency paths of the correlated operators.
 Table \ref{vols}'s second column shows the anomaly
scores for two representative metrics each from $V_1$ and $V_2$.
(Table \ref{vols}'s third column is described later in this section.)
 As expected, none of $V_2$'s metrics are identified as correlated because 
$V_2$ has no contention; while those of $V_1$ are.

\begin{table}[!t]
{
\begin{center}
\begin{tabular}{|c|c|c|}
\hline
Volume,   & Anomaly Score & Anomaly Score \\ 
Perf. Metric              & (no contention in V2) & (contention in V2) \\ \hline\hline
V1, writeIO & 0.894 & 0.894\\ \hline
V1, writeTime & 0.823 & 0.823\\ \hline
V2, writeIO & 0.063 & 0.512\\ \hline
V2, writeTime & 0.479 & 0.879\\ \hline
\end{tabular}
\end{center}
}
\vspace{-5mm}
\caption{Anomaly scores computed during dependency analysis
for performance metrics from Volumes $V_1$, $V_2$}
\label{vols}
\end{table}

\noindent {\bf Module SD:} The symptoms identified so far are: (a) high anomaly scores for operators using $V_1$, (b) high anomaly scores for $V_1$'s performance metrics, and (c) high anomaly score for $O_4$ (only one out of 7 leaf operators using $V_2$).  These symptoms are strong evidence that 
$V_1$'s performance is a cause of the query slowdown,
and $V_2$'s performance is not. Thus, even when
a symptoms database is not available, \dbtool{}
correctly narrows down the search space an administrator has to consider during diagnosis. 
An impact analysis will further point out that the false positive symptom 
due to $O_4$ has little impact on the query slowdown.

Module SD uses a symptoms database that was developed in-house
to diagnose query slowdowns.  $V_1$'s contention due to the SAN misconfiguration
problem was given a high confidence score because all required 
symptoms are found. (The symptoms database had an entry for this
root cause because this problem is very common in production settings.) 
 $V_1$'s contention due to a change in database workload
got a medium confidence score because of a weak correlation 
between the performance of some correlated operators and the rest of the database workload.
All other root cause entries in the symptoms database got low confidence scores.

\noindent{\bf Module IA:} Impact analysis done using the inverse dependency analysis technique
gave an impact score of 99.8\% for the high-confidence root cause
found. This score is high because the slowdown is caused entirely by the contention in $V_1$.

Next, we complicated the problem scenario to test \dbtool{}'s robustness. 
Everything was kept the same except that we created 
extra I/O load on Volume $V_2$ in a bursty manner such that this extra load
had little impact on the query beyond the original impact of $V_1$'s contention. 
Without intrusive tracing, it would not be possible to rule out the extra load on $V_2$ as 
a potential cause of the slowdown. 

Interestingly, \dbtool{}'s integrated approach is still
able to give the right answer. Compared to the previous scenario, there will now be
some extra symptoms due to higher anomaly scores for $V_2$'s performance metrics (as shown in
the third column in Table \ref{vols}). However, root causes with contention-related
symptoms for $V_2$ will still have low confidence because most of the leaf operators
depending on $V_2$ will have low anomaly scores as before. Also, impact scores will 
be low for these causes.

Unlike \dbtool{}, a SAN-only diagnosis tool 
may spot higher I/O loads in both $V_1$ and $V_2$, and attribute 
both of these as potential root causes. Even worse, the tool may 
give more importance to $V_2$ because most of the data is on $V_2$. 
A database-only tool can pinpoint the slowdown in
the operators, but it would likely give several false positives
like a suboptimal buffer pool setting or a suboptimal choice of execution plan.

Some observations from evaluating \dbtool{} on the broad range of scenarios in Table \ref{table:exper} are:

\squishlist

\vspace{1mm}

\item Compared to correlation analysis using advanced models (e.g., Bayesian networks
\cite{cohenosdi04}), KDE can produce accurate results with few tens of samples, and is more
robust to noise in the data. 
\eat{
	For the SAN misconfiguration problem, 
	Figure~\ref{fig:sen} shows the sensitivity of 
	the anomaly scores of three representative operators
	to the number of samples available from the satisfactory runs. 
	Note that the anomaly scores converge within 20 samples.
} 

\vspace{1mm}

\item \dbtool{} can deal with (i) database-level problems whose symptoms propagate to the SAN, 
and vice versa; (ii) independent and concurrent database-level and SAN-level problems;
and (iii) spurious and missing symptoms caused by noise. 

\vspace{1mm}

\item \dbtool{} produces good results even when the symptoms database is 
incomplete. While we expect that entries in the symptoms database are reviewed carefully by administrators,
\dbtool{}'s own modules like correlation, dependency, and impact analysis can be used to identify important symptoms automatically.

\squishend 

\eat{
	\begin{figure}[!t]
	\centering
	\includegraphics[clip=true,width=1\linewidth, height = 1.5in, viewport= 45 0 635 350]{Figures/3operators.eps}
	\caption{Sensitivity of anomaly scores to the number of
	samples. While $O_{22}$ shows highly anomalous behavior, 
	scores for $O_1$ and $O_{11}$ should  be low}
	\label{fig:sen}
	\vspace{-5mm}
	\end{figure}
}

\section{System Operation/Usage}
\label{sec:sysimpl}

\begin{figure*}[t]
	\centering
 	\includegraphics[clip=true,width=1\linewidth,height = 1.5in, viewport = 0 120 575 238 ]{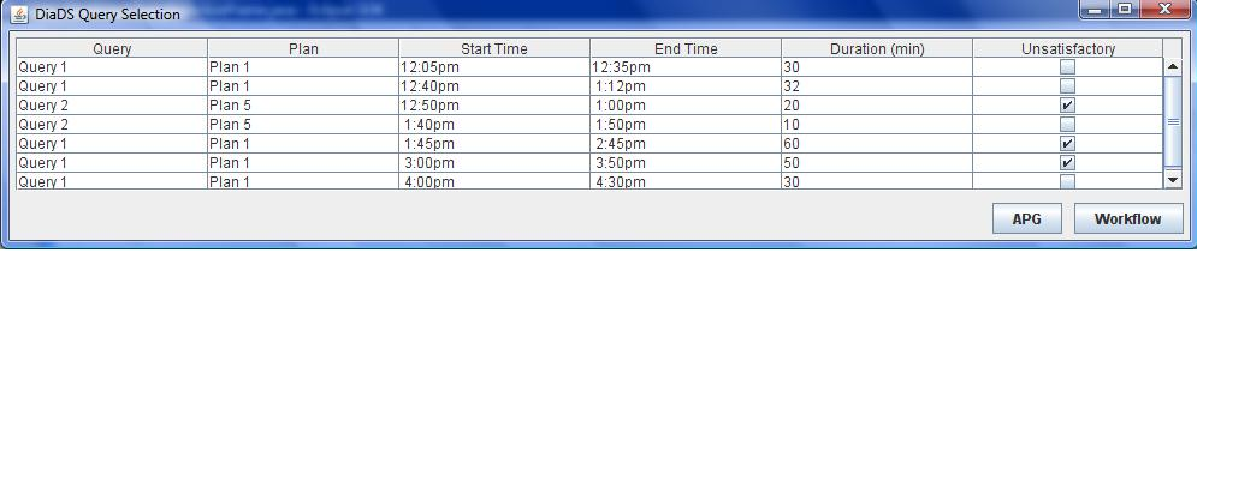}
 	\vspace{-7mm}
	\caption{\dbtool{} query selection screen}
	\label{fig:query_screen}
 	\vspace{-2mm}
\end{figure*}

\begin{figure*}[t]
\centering \includegraphics[clip=true,width=0.7\linewidth]{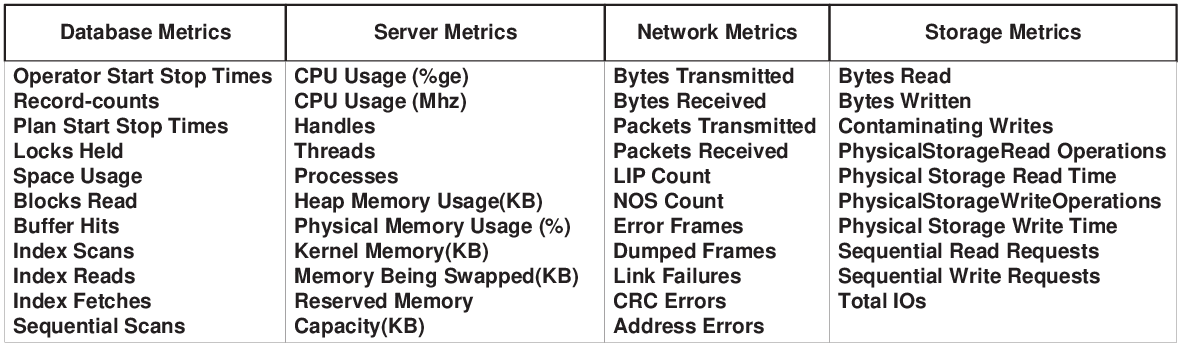}
\vspace{-3mm}
\caption{Performance metrics collected by~\dbtool{}}
\label{metrics-table}
\end{figure*}

\begin{figure}[t]
	\centering
	\includegraphics[clip=true,width=0.7\linewidth, height = 1.5in, viewport= 2 80 340 230]{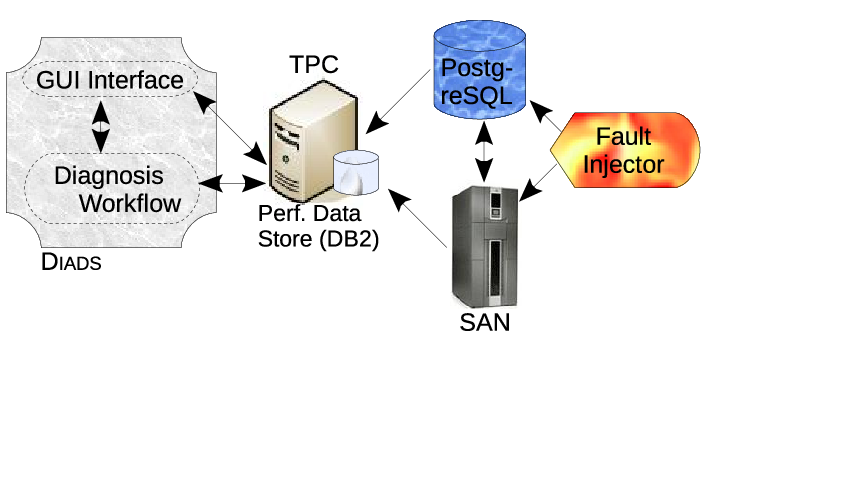}
	\vspace{-3mm}
	\caption{\dbtool{} setup}
	\label{fig:setup}
	\vspace{-5mm}
\end{figure}

Figure \ref{fig:setup} illustrates  an example \dbtool{} deployment as
well  as  the  data  flow   from  the  other  systems  that  \dbtool{}
communicates  with.  This  deployment,   which  we  will  use  in  our
demonstration, consists of:

\squishlist

\vspace{1mm}

\item Data-warehousing queries from the TPC-H benchmark running on a PostgreSQL database server configured to access
data on an enterprise-class IBM storage controller.

\vspace{1mm}

\item The IBM TotalStorage Productivity Center (TPC)~\cite{tpc} running on a separate machine recording configuration
details, statistics, and events from the SAN as well as from PostgreSQL (which was instrumented to report the data to
the management tool). Figure~\ref{metrics-table} shows the key performance metrics collected from the database and SAN.
The monitoring data is stored as time-series data in a DB2 database. 

\vspace{1mm}

\item \dbtool{} running on a separate server: Its graphical user interface (GUI) supports APG-oriented display and
browsing of data collected in the DB2 database. Recall that APGs are views on the monitoring data that combine what DBAs
see -- e.g., data on query plans -- with what SAN administrators see -- data from the numerous SAN components and their
interconnections. APG-oriented visualization was implemented due to comments from administrators that to diagnose a
reasonable fraction of problems all they need to see is the APG diagram. 

\vspace{1mm}
 
\item \dbtool{}'s diagnosis workflow which is invoked on demand: Each module in the workflow is implemented using a
combination of Matlab scripts (for KDE) and Java. \dbtool{} uses a symptoms database that was developed in-house to
handle query slowdowns.

\vspace{1mm}

\item A fault injector\footnote{The fault injector module is used for test purposes and verification of the correctness
of the \dbtool{} results. This module is not required for production deployments.} that can inject a variety of
faults at the database and SAN levels, including SAN misconfiguration, server, disk, or volume contention, RAID
rebuilds, changes in data properties, and table-locking problems. 

\squishend 

\vspace{1mm}

\noindent \dbtool{} diagnosis starts with the administrator identifying a query that has experienced a slowdown. The
diagnosis workflow is then invoked. By default, the workflow is run in a {\em batch} mode where all modules are executed
one after the other, and only the final results are displayed to the administrator.  However, \dbtool{} supports an {\em
interactive} mode where results are displayed after the  completion of each module, and the administrator can edit these results
before they are fed to the next module. In this mode, the administrator  can also re-execute or bypass modules, as well
as stop the execution if the desired result is obtained quickly. In the rest of the section, we describe the interactive
mode and provide explanations on the presented screenshots. 

\dbtool{} first screen (illustrated in Figure \ref{fig:query_screen}) provides a view of the executed queries in the
database.  For each query execution, a corresponding row with the information regarding the execution is presented in the
table. Following information is displayed: 

\squishlist

\vspace{1mm}

\item \textit{Query:}  The actual query string, available as a tool-tip or as a pop-up box when the cell is clicked.

\vspace{1mm}

\item \textit{Plan:}  Executed   plan  for   the  query.   A  visual
representation is showed as a tool-tip or pop-up box on cell click.

\vspace{1mm}

\item \textit{Start time:} The time when the query began execution.

\vspace{1mm}
 
\item \textit{End time:} The time when the query  completed execution.

\vspace{1mm}

\item \textit{Duration:} The number of minutes that the query needed to complete execution.

\vspace{1mm}

\item \textit{Unsatisfactory  check-box:} This check-box is  used by the
administrator to  mark the  query executions that  have unsatisfactory
performance.  \dbtool{} also supports declarative rules for specifying
which query executions are unsatisfactory, e.g., every query execution
that has a running time greater than 30 minutes is unsatisfactory.

\squishend 

\vspace{1mm}

\begin{figure*}[t]
	\centering
	\includegraphics[clip=true,width=1\linewidth,height = 2.5in, viewport = 0 50 575 235 ]{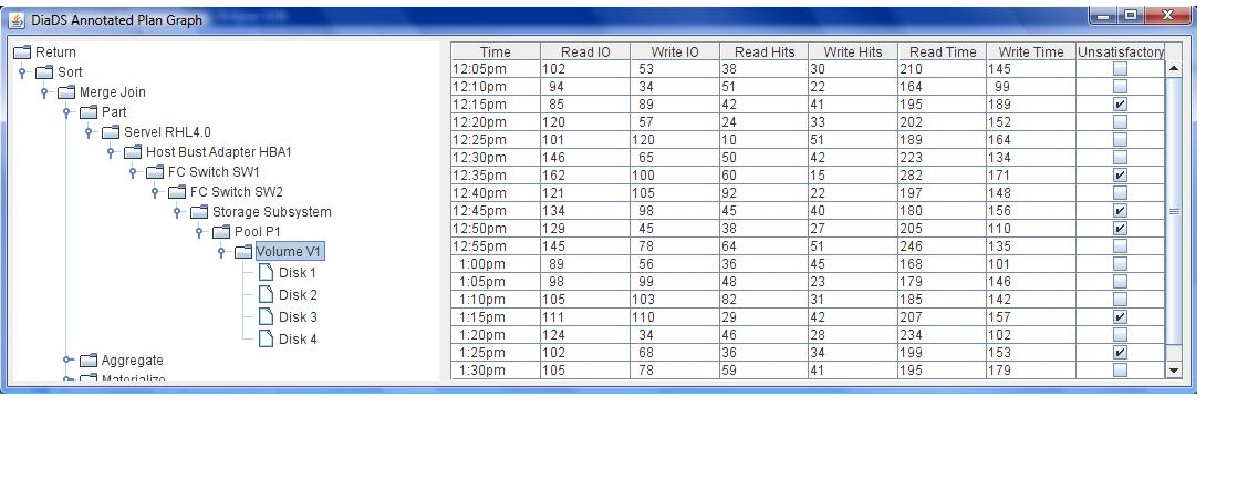}
	\vspace{-7mm}
	\caption{\dbtool{} APG visualization screen}
	\label{fig:apg_screen}
	\vspace{-1mm}
\end{figure*}

\begin{figure*}[t]
	\centering
	\includegraphics[clip=true,width=1\linewidth,height = 2.2in, viewport = 0 35 575 275 ]{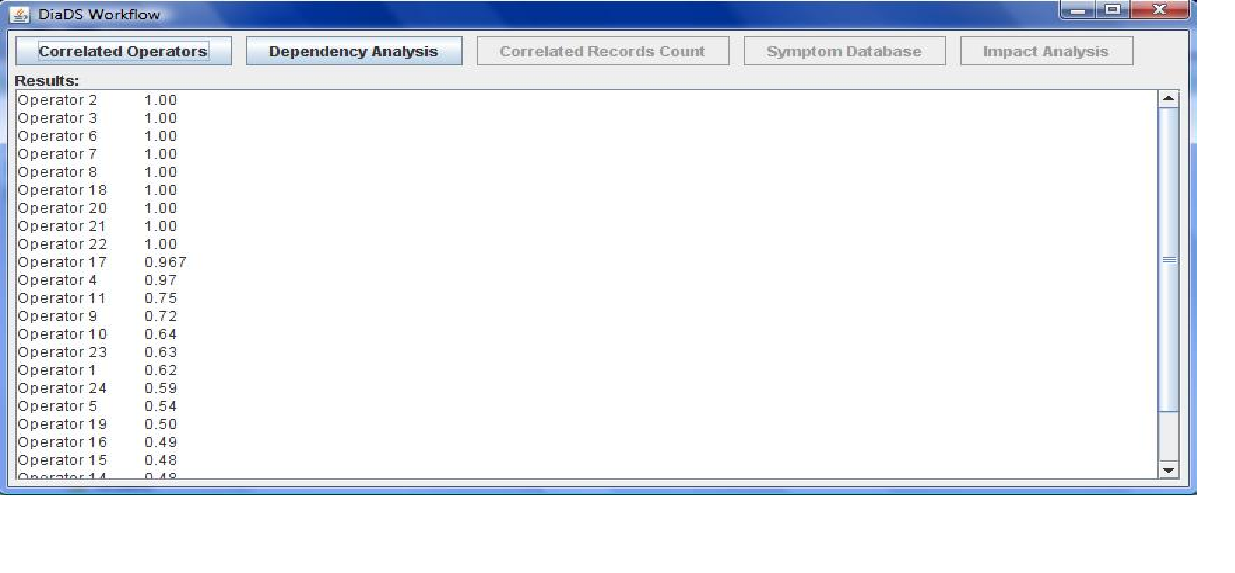}
	\vspace{-7mm}
	\caption{\dbtool{} Interactive workflow execution screen}
	\label{fig:workflow_screen}
	\vspace{-3mm}
\end{figure*}

\noindent The APG button  below the table, navigates the administrator
to the  APG visualization screen  (Figure \ref{fig:apg_screen}), where
the APG for the highlighted  query execution is displayed. On the left
side of this screen, the APG structure is presented as a tree.  Figure
\ref{fig:apg_screen} shows  the path from  Figure \ref{fig:apgs}, that
starts from the  Return operator, goes through the  Index Scan on Part
table and then  all the way to  the disks.  The right side  of the APG
screen contains  a table  of time series  performance metrics  for any
component selected from  the APG.  As each component of  the APG has a
different set  of performance attributes, the  displayed attributes in
the table vary.  However, there  are two attributes that are displayed
for all  components: (i) \textit{Time:}  presenting the time  in which
the  measurement  in  the   corresponding  row  was  taken,  and  (ii)
\textit{Unsatisfactory   check-box:}    presenting   the   performance
categorization of the component.  This check-box is populated based on
the unsatisfactory check-box selection  of the query executions in the
previous  screen.    For  example,  during   an  unsatisfactory  query
execution, all the measurements  taken during the execution are marked
as  unsatisfactory.   However,   the  administrator  can  modify  this
labeling by  selecting this check-box  or not.  As an  example, Figure
\ref{fig:apg_screen}  shows  the  metrics  that  capture  volume  V1's
performance from 12:05pm till 1.30pm.

The  Workflow  button  on  the  query  selection  screen  invokes  the
\dbtool{}        workflow        execution       screen        (Figure
\ref{fig:workflow_screen}). This screen  guides the administrator step
by step  through the  tool workflow. Each  module in \dbtool{}  can be
executed by  clicking on  the corresponding button  on the top  of the
screen. Only  the first execution of  the modules should  be in order,
after that each module can be  re-executed as many times as needed and
in any order.

The result panel  on the screen shows the result  of the last executed
module. Figure \ref{fig:workflow_screen} shows the state of the screen
after  invocation  of  the   module  Correlated  Operators.   As  this
screenshot presents  the first invocation  of the module,  all modules
after dependency analysis are  disabled (i.e., they cannot be executed
next).  When the  last module is executed (Impact  Analysis), the root
causes identified  by \dbtool{}  for the corresponding  query slowdown
are listed in the result panel.


\section{The Potential of Integrated Database and SAN Tools}
\label{sec:potential}

While integrated diagnosis using \dbtool{} solves an important practical
problem, the proposed system and techniques have the potential to enable even
broader functionality. In this section, we present few instances of these
capabilities.

\squishlist

\vspace{1mm}

\item {\it \bfseries What-if analysis}: Often database and storage
administrators have to apply changes within their respective configurations. In
typical enterprises, this either proceeds without regard to impact on the other
layer or requires extensive collaboration between the two teams. In contrast,
using techniques developed in our work, it is easy to conceive an integrated
database and SAN tool that allows administrators to proactively assess the
impact of their planned changes on the other layer. In fact, the impact analysis
component of~\dbtool{} seems to be a promising approach for developing such a
feature. While it may not completely identify all possible problems, it will
serve as a valuable check which can then lead to quicker and more focused
discussions between the teams.

\vspace{1mm}

\item {\it \bfseries Proactive diagnosis and self-healing}: Another useful
extension for~\dbtool{} is to provide proactive diagnosis and importantly,
self-healing capability. The current symptoms database design can be extended to
include, along with symptoms, possible fixes for the root cause of the problem.
Once the tool identifies a root cause, it can then apply the fix to self-heal
the environment. It is important to note that as in real life, the fix may be
required within the database or storage or a combination of both layers. An
integrated approach like ours will be crucial in identifying the right fix and
then applying it in any one layer.

\vspace{1mm}

\item {\it \bfseries Integrated Database and SAN Planning}: Along with
diagnosis, we believe that annotated plan graphs, by capturing information from
the database and SAN layers into a single construct, can lead to smarter
planning and optimization for database deployments over a SAN. For example,
decisions like the choice of storage required for given database workloads or
choice of DB query plan given the storage infrastructure can be intelligently
made using these techniques. An early work by Salem et al~\cite{salem} presented
a similar approach for such integrated planning, though it uses a concatenation
of independent database and storage analysis components. In contrast, annotated
plan graphs provide a much tighter integration with information flow between the
two layers aiding in analysis.

\vspace{1mm}

\item {\it \bfseries Machine Learning and Domain Knowledge Interplay}: One of
the important aspects of our work is the coupling of machine learning and domain
knowledge techniques towards diagnosis. Use of domain knowledge through a
symptoms database serves as a guiding tool to the machine learning algorithms
preventing spurious correlations due to noisy data or event propagation. An
interesting course of future work is to enhance this relationship with machine
learning techniques contributing towards identifying potential symptoms which
can be checked by an expert and added to the symptoms database. Considering that
a symptoms database may never be complete, this provides a self-evolving
mechanism towards bettering the quality of the symptoms databases.

\vspace{1mm}

\item {\it \bfseries Synergy between~\dbtool{} and ADDM~\cite{tune-oracle}}: A
possible deployment of \dbtool{} is along with a more fine-grained diagnosis
tool like Oracle ADDM~\cite{tune-oracle, tune-oracle2} which uses instrumented
code to get operator level timing information. Both use a similar mechanism of
finding symptoms and then mapping them to a root cause. However, our use of
historic performance data helps in answering questions like {\it why did my
query slow down?} while ADDM helps answering questions like {\it why is my query
slow?}. A combination of the tools provides a stronger analysis engine. 
\squishend

\section{Conclusions}
\label{sec:conclusions}

In this paper, we presented  our vision for an integrated database and
SAN  management  framework.   This  framework is  aimed  at  assisting
administrators  in management  tasks that  require an  understanding of
both database and  SAN environments. As an example  of this vision, we
described  a diagnostic  tool,  called \dbtool{},  that supports  root
cause  analysis  for problems  that  span  databases  and SANs.   This
integrated  diagnosis  is based  on  a  novel information  abstraction
called Annotated Plan Graph (APG) that captures the end-to-end mapping
of  database   operators  and   their  dependencies  on   various  SAN
components,  including   performance  and  configuration  information.
Using  a novel  interplay  of machine  learning  and domain  knowledge
(e.g., symptoms  databases), \dbtool{} progressively  drills down from
the  SQL  query  to  execution  plans, operators,  and  eventually  to
performance and  configuration characteristics of  the SAN components.
It  can then  associate impact  of  potential problems  to the  actual
symptoms to identify the root cause of the problem.  We also described
some  experimental scenarios  of  \dbtool{} diagnosis  for root  cause
problems occurring in database and SAN layers.

We  contend  that the  integrated  management  framework  and the  APG
abstraction  presented  in this  paper  enables  a  key capability  in
enterprise data  center management.  By providing visibility  into the
SAN to database  administrators and vice versa, it  allows for smarter
resource planning and improved efficiencies in the data center.

{
\bibliographystyle{abbrv}
\bibliography{bib}
}

\end{document}